\documentclass[paper]{ieice}

\pdfoutput=1

\usepackage[dvips]{graphicx}
\usepackage[varg]{txfonts}
\usepackage{color}
\usepackage{amsfonts,amssymb}
\usepackage{url}
\usepackage{rotating}

\newcommand{\R}{\mathbb{R}}

\newcommand{\PP}{\mathbb{P}}
\newcommand{\ud}{\textrm{d}}

\SpecialSection{Networking Technologies for Cloud Services}
\field{}
\title{}
\title[Probabilistic Resource Management in Cloud]{Dynamic Resource Management in Clouds: A Probabilistic Approach}

\authorlist{%
 \authorentry{Paulo GON\c{C}ALVES}{n}{labelA}
 \authorentry{Shubhabrata ROY}{n}{labelA}
 \authorentry{Thomas BEGIN}{n}{labelA}
  \authorentry{Patrick LOISEAU}{n}{labelB}
}
\affiliate[labelA]{LIP, UMR 5668 Inria - ENS Lyon - UCB Lyon 1 - CNRS. This work was partly funded by the European Project SAIL.}
\affiliate[labelB]{EURECOM, Sophia Antipolis.}

\received{2011}{12}{09}
\revised{2012}{02}{23}

\begin{document}
\maketitle
\begin{summary}
Dynamic resource management has become an active area of research in the Cloud Computing paradigm. Cost of resources varies significantly depending on configuration for using them. Hence efficient management of resources is of prime interest to both Cloud Providers and Cloud Users. In this work we suggest a probabilistic resource provisioning approach that can be exploited as the input of a dynamic resource management scheme. Using a Video on Demand use case to justify our claims, we propose an analytical model inspired from  standard models developed for epidemiology spreading, to represent sudden and intense workload variations. We show that the resulting model verifies a Large Deviation Principle that  statistically characterizes extreme rare events, such as the ones produced by ``buzz/flash crowd effects" that may cause workload overflow in the VoD context.
This analysis provides valuable insight on expectable abnormal behaviors of  systems. We exploit the information obtained using the Large Deviation Principle for the proposed Video on Demand use-case for defining policies (Service Level Agreements). We believe these policies for elastic resource provisioning and usage may be of some interest to all stakeholders in the emerging context of cloud networking
  
\end{summary}

\section{Introduction}
\label{sec:intro}

Users of a Cloud Computing platform can have several numbers of choices regarding server selection (some are compute intensive, some provide better I/O performance, some are superior in networking). Cloud provider such as Amazon offers many different server instances that differ in many aspects with respect to CPU speed, network bandwidth and memory capacity. Each of these instances provides a certain amount of dedicated resource and charges per instance-hour consumed \cite{website:amazon}. A Service Provider finds it to be extremely difficult to optimize the best combination of servers to be deployed in a \emph{Cloud} for his business on a certain application. This problem differs from the concept of traditional distributed computing (like Grid), since the numbers of servers are virtually unlimited  but bandwidth is limited. The choice of deployment of resources can be dynamically tuned using cloud virtualization, that abstracts the IT resources to allow communication and control on-line. Cost of resources varies significantly depending on server types and Cloud Service Providers. 
\newline In most applications, the amount of IT resource that is actually used, is a highly variable quantity that follows the instantaneous activity, and in particular the volume of exchanged traffic when network infrastructures are concerned. Depending on the type of application, the generated workload can be a highly varying process that turns difficult to find an acceptable trade-off between an expensive over-provisioning able to anticipate peak loads and a sub performing resource allocation that does not mobilize enough resources. To bypass this challenge, dynamic bandwidth allocation is an original approach that we chose to investigate in the context of network virtualization. We aim to demonstrate the proof of concept for the case of a Video on Demand (VoD) system by adaptively tuning the provisioned bandwidth to the current application workload. In this paper we have resorted to probabilistic provisioning of resource management; however in some situations it can be used to anticipate resource requirements that can serve as inputs for dynamic resource allocation. \newline
Our work attempts to capture some properties that describe user behaviors or workload generating mechanism of the system and fits them to a mathematical model satisfying particular properties. We leverage these properties to derive a probabilistic assumption on the mean workload of the system at different time resolutions. Embedding the notion of time scale is very important since time scale is by essence intrinsic to dynamicity. In this study we build our system using epidemic models where  Markovian models are widely used and happen to  satisfy to the specific property mentioned above.
\newline
Epidemic information dissemination has been an active area of research in distributed systems, {such as} Peer-to-Peer (P2P) or VoD systems. In \cite{EugEpi2004}, it has been already demonstrated that the epidemic algorithms can be used as an effective solution for information dissemination in the P2P systems as deployed on Internet or ad-hoc networks. {The authors of} \cite{BonalEpistream2008} studied random epidemic strategies like the random peer, latest useful chunk algorithm to achieve optimal information dissemination. However the most relevant work to our study is {derived in} \cite{CarPatMatch2010} where the authors proposed an approach to predict workload for cloud clients. They used auto-scaling algorithm for resource provisioning and validated the result with real-world Cloud client application traces. Our {approach} encompasses both constructive Markovian model to reproduce epidemic information dissemination and workload provisioning aspects. {However, we insist on the fact that its originality stems from the analysis of the Large Deviation property of the proposed Markovian model. The resulting characterization can be viewed as a multi-resolution extension of the classical steady-state distribution for the observable mean value of the random process over different aggregated time scales.}
\newline
After constructing the Markovian mathematical model, we propose two possible and generic ways to exploit these information in the context of probabilistic resource provisioning. They can serve as the input of resource management functionalities of the Cloud environment. It is evident that we can not define elasticity without the notion of a time scale; the Large Deviation Principle (LDP) is capable of automatically integrating the time resolution in automatic description of the system. It is to be noted that Markovian processes do satisfy the LDP, but so do some other models as well. Hence, our proposed probabilistic approach is very generic and can adapt to address any provisioning issues, provided the resource volatility can be resiliently represented by a stochastic process for which the LDP holds true. 
\newline
The rest of the paper is organized as follows. In Section 2 we discuss the VoD system as our use case, followed by a Markovian description of the model in the Section 3. Section 4 presents Large Deviation Principle. We discuss the numerical interpretations in Section 5. Section 6 deals with the probabilistic provisioning scheme, derived from the Large Deviation Spectrum for our use case followed by the conclusion in Section 7. 

\section{Use Case: Video on Demand (VoD)}
\label{sec:usecase}

\begin{figure}[h]
\centering
\includegraphics[width=0.7\columnwidth]{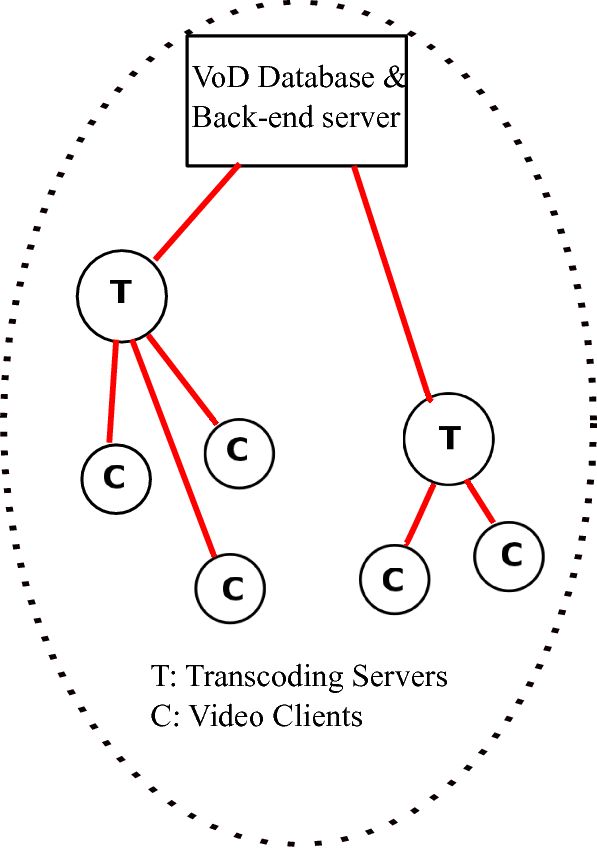}
\caption{\small Basic schematics of a VoD system with transcoding/caching servers}
\label{fig:VoD}
\end{figure}

A VoD service delivers video contents to consumers on request. According to Internet usage trends, users are increasingly getting more involved in the VoD and this enthusiasm is likely to grow. A popular VoD provider like Netflix accounts for around 30 percent of  the peak downstream traffic in the North America and is the ``largest source of Internet traffic overall" \cite{website:sandvine}. In a VoD system, consumers are video clients who are connected to a \textit{Network Provider}. The source video content is managed and distributed by a \textit{Service Provider} from a central data centre. With the evolution of Cloud Computing {and Networking}, the service in a VoD system can be made more scalable by dynamically distributing the caching/transcoding servers across the network providers. Video service providers interact with the network service providers and describe the virtual infrastructures required to implement the service (like the number of servers required, their placements and clustering of resources). The resource provider reserves resource for certain time period and may change it dynamically depending on resource requirement. Such a dynamic approach brings benefits of cost saving in the system through dynamic resource provisioning which is important for service providers as VoD workload is highly variable by nature. However, since {the virtual resources used by Cloud Networking}  have a set-up time which is not negligible, analysis and provisioning of such a system can be very critical from the operators perspective ({\sc capex} versus {\sc opex} trade-off). Figure \ref{fig:VoD} shows a VoD schematic where the back-end server is connected to the data centre and the transcoding (caching) servers are placed across the network providers.

Since VoD has stringent streaming rate requirements, each VoD provider needs to reserve a sufficient amount of server outgoing bandwidth to sustain continuous media delivery. When multiple VoD providers (such as Netflix) are on board to use cloud services from cloud providers, there will be a market between VoD providers and cloud providers, and commodities to be traded in such a market consist of bandwidth reservations, so that VoD streaming performance can be guaranteed.

\begin{figure}[h]
\begin{turn}{90}{\hspace*{8mm}\# number of downloads / day} \end{turn}
\hspace*{-1mm}\includegraphics[width=1\columnwidth]{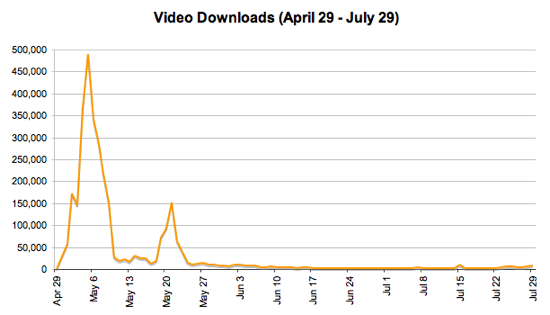}\\
\caption{\small Video server workload: time series displaying a characteristic pattern of flash crowd (buzz effect). Trace obtained from URL: \url{http://waxy.org/2008/05/star_wars_kid_the_data_dump/}}
\label{fig:viral}
\end{figure}

As a buyer in such a market, each VoD provider can periodically make reservations for bandwidth capacity to satisfy its random future demand. A simple way to achieve this is to estimate expectation and variance of its future demand using historical demand information, which can easily be obtained from cloud monitoring services. As an example, Amazon Cloud-Watch provides a free resource monitoring service to Amazon Web Service customers for a given frequency. Based on such estimates of future demand, each VoD provider can individually reserve a sufficient amount of bandwidth to satisfy {\it in average} its random future demand within a reasonable confidence. However, this information is not helpful in case of a ``buzz" or a ``flash crowd" when a video becomes popular very quickly leading to a \emph{flood} of user requests on the VoD servers. Following is one example of ``buzz" where interest over a video``Star Wars Kid" \cite{website:waxy} grew very quickly within a very short timespan. According to \cite{website:bbc} it was viewed more than 900 million times within a short interval of time making it one of the top viral videos. Figure \ref{fig:viral} plots the original server logs for the Star Wars Kid debacle \cite{website:waxy}.

In situations like the one described in Figure \ref{fig:viral}, variance estimation or more generally steady state distribution can not explain burstiness of such event as time resolution is excluded from the description. The LDP, by virtue of its multi-resolution extension of the classical steady-state distribution, can describe the dynamics of rare events like this, which we believe can be of some interest for the VoD service providers.

\section{Markov Model to describe the behavior of the users}
\label{sec:markov}
{Epidemic models commonly subdivide a population into several compartments:} susceptible (noted $S$) to designate the persons who can get infected, and contagious (noted $C$) for the persons who have contracted the disease. This contagious class can further be categorized into two parts: the infected subclass ($I$) corresponding to the persons who are currently suffering from the disease and can spread it, and the recovered class ($R$) for those who got cured and do not spread the disease anymore \cite{BarDynaPro2008}. There can be more categories {that fall outside} the scope  of our current work.  In these models $(N_{S}(t))_{t \geq 0}$, $(N_{I}(t))_{t \geq 0}$ and $(N_{R}(t))_{t \geq 0}$ are stochastic processes representing the {time} evolution of susceptible, infected and recovered populations respectively.
\newline {Similarly,} information dissemination in a social {network} can be viewed as an epidemic spreading {(through gossip)}, where the ``buzz"  is a special event where interest {for} some {particular} information increases {drastically} within a very short period {of time}. Following {the lines of} related works, we claim that the above mentioned epidemic models can appropriately be adapted to represent the way information spreads among the users in a VoD system.  In the case of a VoD system, infected $I$ refers to the people who  are currently watching the video and can spread the information about it. {In our setting,} $I$ directly represents the current workload which is the current aggregated video requests from the users. Here, we consider the workload as the total number of current viewers, but it can also refer to total bandwidth requested at the moment. The class $R$  refers to the past viewers. {In contrast to the classical epidemic case, we introduce a memory effect in our model, assuming that the $R$ compartment can still propagate the gossip during a certain random latency period.}
{Then, we define the probability within a small time interval ${\rm d}t$, for a susceptible individual to turn into an active viewer, as follows}:
\begin{equation}
\mathbb{P}_{S\to C} = {(l+ (N_I(t)+N_R(t)) \, \beta) {\rm d}t} + o({\rm d}t)
\label{eq:transition-prob}
\end{equation}
where $\beta > 0$ is the rate of information dissemination per unit time and $l>0$ fixes the rate of spontaneous viewers.
The  instantaneous rate of newly active viewers in the system at time $t$ is thus:
\begin{equation}
\lambda(t) = l+ (N_I(t)+N_R(t))\beta.
\label{eq:infinity}
\end{equation}
Equation (\ref{eq:infinity}) corresponds to the arrival rate $\lambda(t)$ of a non-homogeneous (state dependant) Poisson process. This rate varies linearly with $N_I(t)$ and $N_R(t)$.
%

\begin{figure} [h]
\centering
\includegraphics[width=0.7\columnwidth]{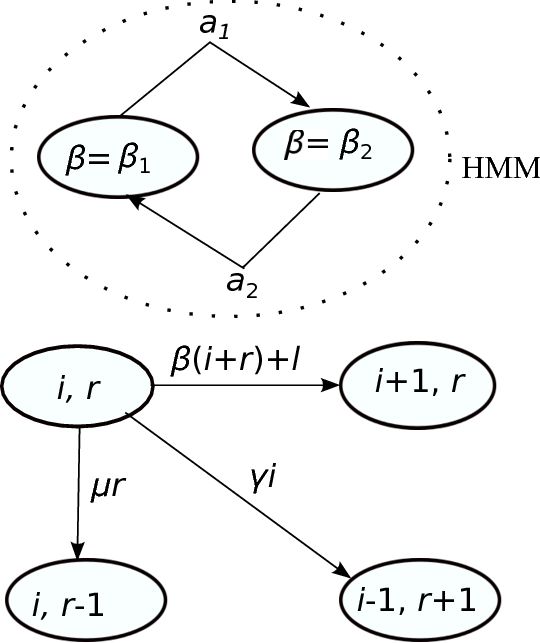}
\caption{\small Markov chain diagram representing the evolution of the Current viewers ($i$) and Past Viewers ($r$) populations with a Hidden Markov Model. }
\label{fig:Markov}
\end{figure}

To complete our model we assume that the watch time of a video is exponentially distributed with rate $\gamma$. As already mentioned, it also deems reasonable to consider that a past viewer will not keep propagating the gossip about a video indefinitely, but  remains active only for a latency random period that we also assume exponentially distributed  with rate $\mu$ (in general $\mu \ll \gamma$). Another important consideration of the model is the maximum allowable viewers ($I_{\textrm{max}}$) at any instant of time. This assumption conforms to the fact that the resources in the system are physically limited. For the sake of numerical tractability and without loss of generality, we also assume the number of past (but spreading rumour) viewers at a given instant to be bounded by a maximum value ($R_{\textrm{max}}$). With these assumptions, and posing ($N_I(t)=i,N_R(t)=r)$ the current state of the Markov processes, the probability that the process reaches a different state $(i' < I_{\textrm{max}},r'< R_{\textrm{max}})$ at time $t+{\rm d}t$  (${\rm d}t$ being small) reads:
\begin{eqnarray}
\lefteqn{\mathbb{P}(i', r' | i, r) } \\
&=& (l+(i+r) \beta){\rm d}t + o({\rm d}t) \hspace*{5mm} \mbox{for } (i'=i+1,r'=r), \nonumber \\
&=& (\gamma i){\rm d}t + o({\rm d}t)  \hspace*{12mm}\mbox{for } (r'=r+1,i'=i-1), \nonumber \\
&=& (\mu r){\rm d}t + o({\rm d}t) \hspace*{17mm} \mbox{for } (r'=r-1,i'=i), \nonumber \\
&=& o({\rm d}t)  \hspace*{45mm} \mbox{otherwise.} \nonumber
 \label{eq:MC2-transition}
\end{eqnarray}
This process defining the evolution of the current viewer and past viewer populations  is a finite and irreducible Markov chain. It is to be noted that $l  > 0$ precludes the process to reach an absorbing state. This chain is ergodic and admits a stationary regime. \newline
Above mentioned descriptions define the mechanism of information dissemination in the community in normal situations. A buzz event differs from this situation by a sudden increase of the dissemination rate $\beta$. In order to adapt the model to buzz we resort to Hidden Markov Model (HMM) to be able to reproduce the change in $\beta$. Without loss of generality we consider only two states. One with dissemination rate $\beta=\beta_{1}$ corresponds to the buzz-free case described above, and another hidden state corresponding to the buzz situation, where the value of $\beta$ increases significantly and takes on a value  $\beta_{2} \gg \beta_{1}$. Transitions between these two hidden and memoryless Markov states occur  with rates $a_{1}$ and $a_{2}$ respectively (see Figure \ref{fig:Markov}). These rates characterize the buzz in terms of frequency, magnitude and duration. 

\section{Large Deviation Principle}
\label{sec:LDP}

Consider a continuous-time Markov process $(X_t)_{t\ge0}$, taking values in a finite state space $S$, of rate matrix $A = (A_{ij})_{i\in S, j\in S}$.  In our case $X$ is a vectorial process $X(t) = \left( N_I(t), N_R(t)\right), \forall t\ge0$, and $S = \{ 0, \cdots, I_{\textrm{max}} \} \times \{ 0, \cdots, R_{\textrm{max}} \}$. If the rate matrix $A$ is irreducible, then the process $X$ admits a unique steady-state distribution $\pi$ satisfying $\pi A = 0$. Moreover, by Birkhoff ergodic theorem, it is known that for any mapping $\Phi: S \to \R$, the sample mean of $\Phi(X)$ at scale $\tau$, i.e. $1/\tau\cdot\int_{0}^{\tau} \Phi(X_s) \ud s$  converges almost-surely towards the mean of $\Phi(X)$ under the steady-state distribution, as $\tau$ tends to infinity. The function $\Phi$ is often called the \emph{observable}. In our case, as we are interested in the variations of the current number of users $N_I(t)$, $\Phi$ will simply be the function that selects the first component: $\Phi(N_I(t), N_R(t)) = N_I(t)$. 
The large deviations principle (LDP), which holds for irreducible Markov processes on a finite state space \cite{Varadhan08a}, gives a efficient way to estimate the probability for the sample mean calculated over a large period of time $\tau$ to be around a value $\alpha\in\R$ that deviates from the almost-sure mean:
\begin{equation}
\hspace*{-6mm} \lim_{\epsilon\to0} \lim_{\tau\to\infty} \frac{1}{\tau} \log \PP \left\{  \int_{0}^{\tau} \Phi(X_s) \ud s \in [\alpha - \epsilon, \alpha+\epsilon]  \right\} = f(\alpha).
\label{eq:LD-proba}
\end{equation}
The mapping $\alpha\mapsto f(\alpha)$ is called the large deviations spectrum (or the rate function). For a given function $\Phi$, it is possible to compute the theoretical large deviations spectrum from the rate matrix $A$ as follows. One first computes, for each values of $q\in\R$, the quantity $\Lambda (q)$ defined as the principal eigenvalue (\emph{i.e.,} the largest) of the matrix with elements $A_{ij}+q \delta_{ij}\Phi(j)$ ($\delta_{ij}=1$ if $i=j$ and 0 otherwise). Then the large deviations spectrum can be computed as the Legendre transform of $\Lambda$: 
\begin{equation}
f(\alpha) = \sup_{q\in\R} \left\{q \alpha - \Lambda (q) \right\}, \forall \alpha\in\R.
\label{eq:sup}
\end{equation}

As described in Equation(\ref{eq:LD-proba}), $\alpha_{\tau}=\langle i \rangle_{\tau}$  corresponds in our study case, to the mean number of users $i$ observable over a period of time of length $\tau$ and $f(\alpha)$ relates to the probability of its occurrence as follows:
\begin{equation}
\mathbb{P} \{\langle i \rangle_{\tau} \approx \alpha \} \sim e^{\tau \cdot f(\alpha)}.
\label{eq:joint-prob}
\end{equation}

Interestingly also, if the process is strictly stationary (\emph{i.e.} the initial distribution is invariant) the same large deviation spectrum $f(\cdot)$ can be estimated from a single trace, provided that it is "long enough'' \cite{Barral11a}. We proceed as follows: At a scale $\tau$, the trace is chopped into $k_{\tau}$ intervals $\{I_{j,\tau}=[(j-1)\tau, j\tau[,\,j=1,\ldots,k_{\tau}\}$ of length $\tau$ and we have (almost-surely), for all $\alpha\in\R$:
\begin{equation}
\hspace*{-7mm}
\begin{array}{c}
\displaystyle{f_{\tau}(\alpha, \epsilon_{\tau}) = \frac{1}{\tau} \log \frac{ \#\left\{ j:   \int_{I_{j,\tau}} \Phi(X_s) \ud s \in [\alpha - \epsilon_{\tau}, \alpha+\epsilon_{\tau}]  \right\} }{k_{\tau}} } \\[4mm]
 \mbox{and }\displaystyle{ \lim_{\tau\to\infty} f_{\tau}(\alpha, \epsilon_{\tau}) =  f(\alpha).}
 \end{array}
\label{eq:falpha}
\end{equation}


In practice, for the empirical estimation of the large deviations spectrum, we use a similar  estimator as the one derived in \cite{Barral11b} and also used in \cite{Loiseau10a}.  At scale $\tau$, we compute for each $q\in \R$ the values of $\Lambda_{\tau}^{\prime} (q)$ and $\Lambda_{\tau}^{\prime\prime} (q)$, where 
{$\Lambda_{\tau} (q) = \tau^{-1} \log \left(k_{\tau}^{-1}\sum_{j=1}^{k_t} \exp \left( q \int_{I_{j,\tau}}  \Phi(X_s) \ud s  \right)\right)$}. Then, for each value of $\tau$, we count the number of intervals $I_{j, \tau}$ verifying the condition in expression (\ref{eq:falpha}) and estimate the scale-dependant empirical {\em log-pdf} $f_{\tau}(\alpha, \epsilon_{\tau})$, with the  adaptive choices derived in \cite{Barral11b}:
\begin{equation}
\alpha_{\tau}  = \Lambda_{\tau}^{\prime} (q) \quad \textrm{ and }  \quad \epsilon_{\tau} = \sqrt{ \frac{-\Lambda_{\tau}^{\prime\prime} (q)}{\tau} }.
\label{eq:alpha}
\end{equation}
Let us now illustrate the LDP in the context of the specific VoD use case, where $X$ would correspond to $(i, r)$, the bi-variate Markov process. $\Phi(X)$ is $i$, the observable and
$\int^\tau_0 \Phi(X_{S})\,ds = \langle i \rangle_{\tau}$ corresponds to the average number of users with a period $\tau$.

\section{Numerical Interpretations}
\label{sec:validation}

\begin{figure}[h]
\begin{center}
\begin{tabular}{c}
{\normalsize (a)} \\
\hspace*{-86mm}\includegraphics[width=1\columnwidth]{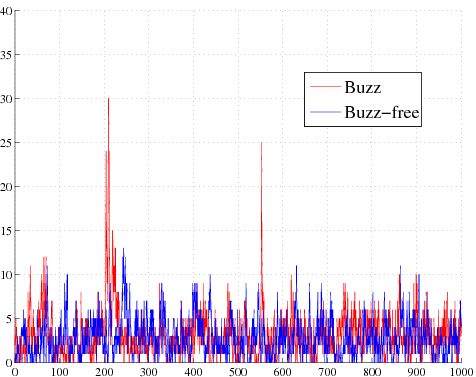} 
\hspace*{-82mm}\begin{turn}{90}{\hspace*{22mm}\# of current viewers}\end{turn}\\
Time (hours)\\[2mm]
{\normalsize (b)} \\
\hspace*{-86mm}\includegraphics[width=1\columnwidth]{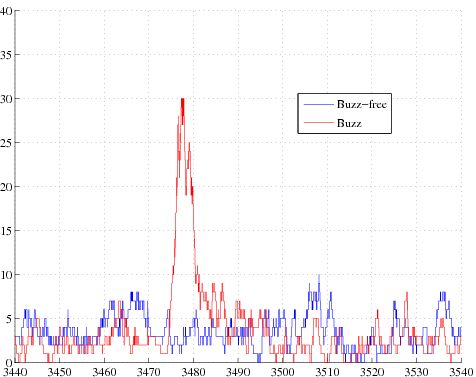} 
\hspace*{-82mm}\begin{turn}{90}{\hspace*{22mm}\# of current viewers}\end{turn}\\
Time (hours)\\
\end{tabular}
\end{center}
\caption{{\small Plot (a): Workload $N_I(t)$ generated according to the model depicted in Figure \ref{fig:Markov}  (For the buzz case: $\beta_{1} = 0.1$, $\beta_{2} = 0.8$, $\gamma = 0.7$, $\mu = 0.3$, $l = 1.0$, $a_{1}=0.006$ and $a_{2}=0.6$. For the buzz-free case: $\beta_1=\beta_2=\beta= 0.1$, $\gamma = 0.7$, $\mu = 0.3$, $l = 1.0$). In both cases, $I_{\rm max} = 30, R_{\rm max}=60$.
Plot (b): Zoomed in view of a buzz event.}}
\label{fig:traces}
\end{figure}

We simulate the proposed workload model and generate two time series corresponding to the buzz and to the buzz free situations. We developed our simulator in \emph{C} programming environment, by creating several parallel child processes (client) that communicate with a parent process (server) to disseminate information. The child process is in any of the susceptible, active viewers or past viewers states at a particular instant of time. When it is in the past viewers state it randomly chooses another process (using process id) and communicates with the parent to infect him. The parent process maintains a table with the status (which state a process is in) of each process. If the chosen process is not already in active viewers or in past viewers states it gets infected. We have chosen UDP socket-pairs in order to facilitate communication between the processes. For fair and consistent comparisons, we carefully tuned the values of the model parameters so as to obtain the same mean workload for both resulting traces. %
In Figure \ref{fig:traces}(a) the bursty transients represent the buzz effect. It reflects sudden and sharp increases of workload due to intense dissemination of popular videos. The zoomed in view displayed in Figure \ref{fig:traces}(b) shows the characteristic pattern of a buzzy transient, that is to say a sharp increase ($\beta_{1} \to \beta_{2}$) and a slow decrease (owing to $\beta_{2} \to \beta_{1}$ and memory effect of the model).

\begin{figure} [h]
\begin{center}
\begin{tabular}{cc}
\begin{turn}{90}{\hspace*{26mm}$\mathbb{P}(i)$}\end{turn} &
\hspace*{-2mm}\includegraphics[width=.9\columnwidth]{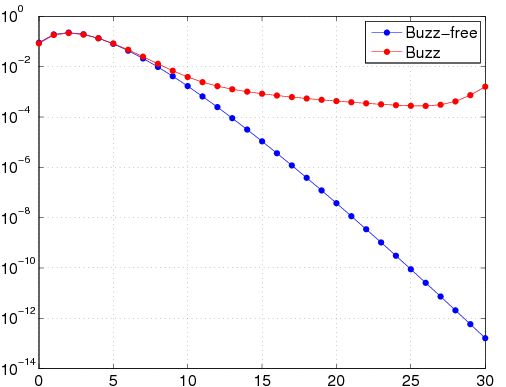} \\
& $i$ (number of current viewers)
\end{tabular}
\end{center}
\caption{\small Steady-state probabilities for the number of current viewers with buzz and buzz-free scenarios (Y-axis in log-scale).}
\label{fig:ss}
\end{figure}

This clear evidence of our model's ability to capturing the buzz effect is moreover confirmed by the numerical steady-state distributions $\mathbb{P}(i)$ displayed in Figure \ref{fig:ss}. As compared to the buzz-free case, the buzz distribution presents a thicker tail indicating that the instantaneous workload $i$ takes on larger values with higher probability. To include the notion of time scale in the results one needs to consider along with the steady-state distribution the time coherence of the underlying process, viz. it's covariance structure.
However, except for the trivial case of uncorrelated processes deriving the statistics of the local average process at any resolution is a hard problem in general.  

\begin{figure*}[t]
\begin{center}
\begin{tabular}{ccc}
{\normalsize (a)} & \hspace*{-8mm} {\normalsize (b)} & \hspace*{-8mm} {\normalsize (c)} \\
\hspace*{-10mm}\includegraphics[width=0.70\columnwidth]{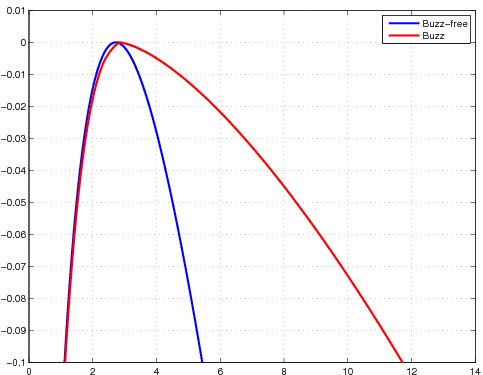}
&
\hspace*{-0mm}\includegraphics[width=0.70\columnwidth]{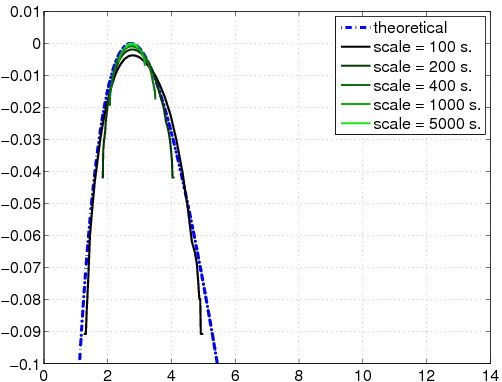} 
&
\hspace*{-0mm}\includegraphics[width=0.70\columnwidth]{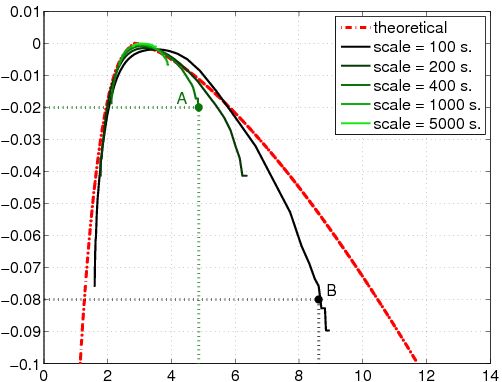}\\[-26mm]
\hspace*{-72mm}\begin{turn}{90}$f(\alpha)$\end{turn} 
& 
\hspace*{-61mm}\begin{turn}{90}$f(\alpha)$\end{turn} 
&
\hspace*{-61mm}\begin{turn}{90}$f(\alpha)$\end{turn}  \\[20mm]
$\alpha = \langle i \rangle_{\tau}$ & $\alpha = \langle i \rangle_{\tau}$ & $\alpha = \langle i \rangle_{\tau}$  \\
\end{tabular}
\end{center}
\caption{\small Large Deviations spectra corresponding to the traces of Figure \ref{fig:traces}. (a) Theoretical spectra for the buzz free (blue) and for the buzz (red) scenarii. (b) \& (c) Empirical estimations of $f(\alpha)$ at different scales from the buzz free and the buzz traces, respectively. }
\label{fig:LDP}
\end{figure*}

Intrinsically, Large Deviation Principle naturally embeds this time scale notion into the statistical description of the aggregated observable at different time resolutions. As expected, the theoretical LD spectra displayed in Figure~\ref{fig:LDP}(a) reach their maximum for the same mean number of users. This apex is the almost sure value as described in Section~\ref{sec:LDP}. As the name suggests almost sure workload ($\alpha_{a.s}$) corresponds to the mean value that we almost surely observe on the trace. More interestingly though, the LD spectrum corresponding to the buzz case, spans over a much larger interval of observable mean workloads than that of the buzz-free case. This remarkable support widening of the theoretical spectrum shows that LDP can accurately quantify the occurrence of extreme, yet rare events.

Plots (b)-(c) of Figure \ref{fig:LDP}  compare theoretical and empirical large deviation spectra obtained for the two traces. For each given scale ($\tau$) the empirical estimation procedure yields one LD estimate. These empirical estimates at different scales superimpose for a given range of $\alpha$. This is reminiscent of the scale invariant property underlying the large deviation principle. If we focus on the supports of the different estimated spectra, the larger the time scale $\tau$ is, the smaller becomes the interval of observable value of $\alpha$. This is coherent with the fact that for a finite trace-length the probability to observe a number of current viewers, that in average, deviates from the nominal value ($\alpha_{a.s}$) during a period of time ($\tau$) decreases exponentially fast with $\tau$. To fix the ideas, the estimates of plot (c),  indicate that  for a time scale $\tau =400\,sec.$, the maximum observable mean number of users is around 5 with probability $2^{400\cdot(-0.02)} \approx 35.10^{-5}$ (point A), while it increases up to $9$ with the same probability ($2^{100\cdot(-0.08)}$) for $\tau =100\,sec.$ (point B).

\section{Probabilistic Provisioning}
\label{sec:provision}

Retuning to our VoD use case, we now sketch two possible schemes for exploiting the Large Deviation description of the system to dynamically provision the allocated resources:

\begin{itemize}
\item {\it Identification of the reactive time scale for reconfiguration}: Find a relevant time scale that realizes a good trade-off between the expectable level of overflow associated to this scale and a sustainable {\sc opex} cost induced by the resources reconfiguration needed to cope with the corresponding flash crowd.
\item {\it Link capacity dimensioning}: Considering a maximum admissible loss probability, find the  safety margin  that it is necessary to provision on the link capacity, to guarantee the corresponding QoS. 
\end{itemize}

\subsection{Identification of the reactive time scale for reconfiguration}
\label{sec:scale}

We consider the case of a VoD service provider who wants to determine the reactivity scale at which it needs to reconfigure its resource allocation. This quantity should clearly derive from a good compromise between the level of congestion (or losses) it is ready to undergo, i.e. a tolerable performance  degradation, and the price it is willing to pay for a frequent reconfiguration of its infrastructure. Let us then assume that the VoD provider has fixed admissible bounds for these two competing factors, having determined the following quantities:
\begin{itemize}
\item $\alpha^* > \alpha_{\rm a.s.}$: the deviation threshold  beyond which it becomes worth (or mandatory) considering to reconfigure the resource allocation. This choice is uniquely determined by a {\sc capex} performance concern. 
\item $\sigma^*$: an acceptable probability of occurrence of these overflows. This choice is essentially guided by the corresponding {\sc opex} cost.
\end{itemize}

Let us moreover suppose, that the LD spectrum $f(\alpha)$ of the workload process was previously estimated, either by identifying the parameters of the Markov model used to describe the application, or empirically from collected traces. Then, recalling the probabilistic interpretation we surmised in relation (\ref{eq:joint-prob}), the minimum reconfiguration time scale $\tau^*$ for dynamic resource allocation, that verifies the sought compromise, is simply the solution of the following inequality:
\begin{equation}
\tau^* = \max\{\tau : \mathbb{P}{\{ \langle i \rangle_{\tau} \geq \alpha^* \} }  =  \int^\infty_{\alpha^*} e^{\tau f_{\tau}(\alpha)}\,d{\alpha}  \geq \sigma^*\},
\label{eq:meanusr}
\end{equation}
with $f_{\tau}(\alpha)$ as defined in expression (\ref{eq:falpha}).

From a more general perspective though, we can see this problem as an underdetermined system involving 3 unknowns ($\alpha^*$,$ \tau^*$ and $\sigma^*$) and only one relation (\ref{eq:meanusr}). Therefore, and depending on the sought objectives, we can imagine to fix any other two of these variables and to determine the resulting third so that it abides with the same inequality as in expression (\ref{eq:meanusr}).

\subsection{Link capacity dimensioning}
\label{sec:serv}

We now consider an architecture dimensioning problem from  the infrastructure provider perspective. Let us assume that the infrastructure and the service providers have come to a Service Level Agreement (SLA), which among  other things, fixes a tolerable level of losses due to link congestion.  We start considering the case of a single VoD server and address the following question: What is the minimum link capacity $C$ that has to be provisioned  such that we meet the negotiated QoS in terms of loss probability? Like in the previous case, we assume that the estimated LD spectrum $f(\alpha)$ characterizing the application has been priorly identified. A rudimentary SLA would be to  guarantee a loss free transmission for  the {\em normal} traffic load only: this loose QoS would simply amount to fix $C$ to the almost sure workload $\alpha_{\rm a.s.}$. Naturally then, any load overflow beyond this value will result in goodput limitation (or losses, if there is no buffer to smooth out exceeding loads). For a more demanding QoS, we are led to determine the necessary safety margin $C_0>0$ one has to provision above $\alpha_{\rm a.s.}$ to absorb the exact amount of overruns corresponding to the loss probability $p_{\rm loss}$ that was negotiated in the SLA.  From the interpretation of the large deviation spectrum provided in Section \ref{sec:LDP},  this margin $C_0$ is determined by the resolution of the following inequality:
\begin{eqnarray}
C_0 & \:{\rm :} &  \int_{\alpha_{\rm a.s.}+C_0}^{\infty} \int_{\tau_{min}}^{\tau_{max}} e^{\tau\cdot f(\alpha)}\,{\rm d}\tau\,{\rm d}\alpha~ \leq ~p_{\rm loss} \nonumber \\\label{eq:ploss-1}
& \:{\rm :} &\int_{\alpha_{\rm a.s.}+C_0}^{\infty} \frac{e^{\tau_{max}\cdot f(\alpha)}-e^{\tau_{min}\cdot f(\alpha)}}{f(\alpha)} \,{\rm d}\alpha ~\leq ~p_{\rm loss}
\end{eqnarray}
In this expression,  $\tau_{min}$ is typically  determined by the size $Q$ of the buffers that is usually provisioned to dampen the traffic volatility. In that case,
\begin{equation}
\tau_{\rm min} = \frac{Q}{\alpha-(\alpha_{a.s.}+C_0)},
\label{eq:tau_min}
\end{equation} 
corresponds to the maximum burst duration that can be buffered without causing any loss at rate $\alpha>C=\alpha_{a.s.}+C_0$.  
As for $\tau_{\rm max}$, it relates to the maximum period  of reservation dedicated to the application. Most often though, the characteristic time scale of the application exceeds the dynamic scale of flash crowds  by several orders of magnitude, and $\tau_{\rm max}$ can then simply be set to infinity. With these particular integration bounds, Equation (\ref{eq:ploss-1}) simplifies to
\begin{equation}
\begin{array}{c}
\displaystyle{C_0 = C-\alpha_{a.s.} \:{\rm :} \int_{C}^{\infty} \frac{-1} {f(\alpha)}\,e^{\frac{Q}{\alpha-C} f(\alpha)}\,{\rm d}\alpha ~ \leq ~ p_{\rm loss}},
\\[5mm]
\end{array}
\label{eq:ploss-2}
\end{equation}
a decreasing function of $C$, which can be solved using a simple bisection technique. \\
As long as the server workload remains below $C$,  this resource dimensioning  guarantees that no loss occurs. All overrun above this value will produce losses, but we ensure that the frequency (probability) and duration of these overruns are such that the loss rate remains  conformed to the SLA. 
The proposed approach clearly contrasts  with resource over-provisioning  that does not seek at optimizing the {\sc capex} to comply with the loss probability tolerated in the SLA.

The same provisioning scheme can straightforwardly be generalized to the case of several applications sharing a common set of resources. To fix the idea, let us consider an infrastructure provider that wants to  host $K$ VoD servers over the same shared link. A corollary question is then to determine how many servers $K$ can the fixed link capacity $C$ support, while guaranteeing  a prescribed level of losses. If the servers are independent, the probability for two of them to undergo a flash crowd simultaneously is negligible. For ease and without loss of generality, we moreover suppose that they are identically distributed and modeled by the same LD spectrum $f^{(k)}(\alpha)=f(\alpha)$ with the same nominal workload $\alpha^{(k)}_{\rm a.s.}=\alpha_{\rm a.s.},\,k=1,\ldots K$. 
Then, following the same reasoning as in the previous case of a single server, the maximum number $K$ of servers reads:

\begin{equation}
K = \mbox{arg}\max_{K} \left( C -K\cdot \alpha_{\rm a.s.}\right) \leq C_0,
\end{equation}
where the safety margin $C_0$ is defined as in expression (\ref{eq:ploss-2}).

Then, depending on the agreed {\it Service Level Agreements}, the infrastructure provider can easily offer different levels of probability losses (QoS) to its VoD clients, and adapt  the number of hosted servers, accordingly.

\begin{figure} [h]
\begin{tabular}{c}
\includegraphics[width=0.9\columnwidth]{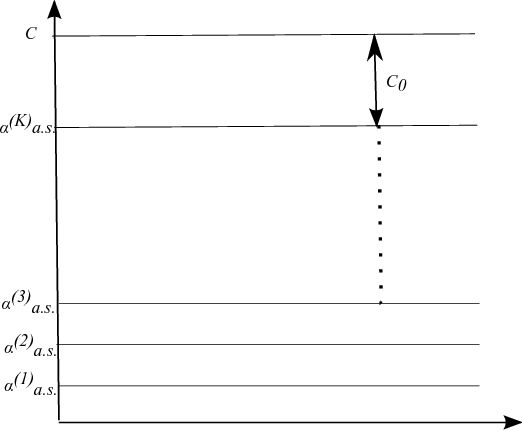}\\
{\normalsize Time}
\end{tabular}
\caption{\small Dimensioning $K$, the number of hosted servers sharing a fixed capacity link $C$. The safety margin $C_0$ is determined according to the probabilistic loss rate negotiated in the {\it Service Level Agreement} between the infrastructure provider and the VoD service provider.}
\label{fig:caplan}
\end{figure}

\section{Conclusion}
\label{sec:conclusion}

The objective of this work is to harness probabilistic methods for resource provisioning in the Clouds. We illustrate our purpose with a Video on Demand scenario, a characteristic service whose demand relies on information spreading. Adopting a constructive approach to capture the users' behavior, we  proposed a simple, concise and versatile model for generating the workload variations in such context. 
A key-point of this model is that it permits to reproduce the workload time series with a Markovian process, which is known to verify a Large Deviation Principle (LDP). This particularly interesting property yields a large deviation spectrum whose interpretation enriches the information conveyed by the standard steady state distribution: For a given observation (workload trace), LDP allows to infer (theoretically and empirically) the probability that the time average workload, calculated at an arbitrary aggregation scale, deviates from its nominal value (i.e. almost sure value). 
\newline
We leveraged this multiresolution probabilistic description to conceptualize two different management schemes for dynamic resource provisioning. As explained, the rationale is to use large deviation information to help network and service providers together to agree on the best {\sc capex}-{\sc opex} trade-off. Two major stakes of this negotiation are: {\it (i)} to determine the largest reconfiguration time scale adapted to the workload elasticity and {\it (ii)} to dimension VoD server so as to guarantee with upmost probability the Quality of Service imposed by the negotiated  Service Level Agreement.
\newline
More generally though, the same LDP based concepts can benefit any other ``Service on Demand" scenarii to be deployed on dynamic cloud environments. 

\bibliographystyle{ieicetr}
\bibliography{refs}

\profile{Paulo Gon\c{c}alves}{%
graduated from the Signal Processing
Department of ICPI Lyon (now CPE Lyon), France in 1993. He
received the Masters (DEA) and Ph.D. degrees in signal processing
from the Institut National Polytechnique de Grenoble,
France, in 1990 and 1993 respectively. While working toward his
Ph.D. degree, he was with Ecole Normale Sup\'erieure de Lyon
(ENS-Lyon). Since 1996, he is associate researcher at
Institut National de Recherche en Informatique et
Automatique (Inria). He is currently head of the Inria team ``RESO" at the
Laboratoire de l'Informatique du Parall\'elisme (LIP) of ENS-Lyon.
P. Gon\c{c}alves research interests are in multiscale analysis 
(signals, images and systems) and in wavelet-based statistical
inference. His principal application is in metrology and deals
with grid traffic statistical characterization and modelling for
protocol quality assessment and control.
}

\profile{Shubhabrata Roy}{%
did his Bachelors in Electrical Engineering from the Jadavpur University, India and Masters in Communication and Networks at SSSUP in CNR, Pisa. He is 
currently pursuing his PhD at Ecole Normale Sup\'erieure de Lyon under the supervision of Paulo Gon\c{c}alves and Thomas Begin. His research interests include Network Virtualization, Cloud Computing and Stochastic Processes.
}

\profile{Thomas Begin}{%
is an Assistant Professor at Université Claude Bernard Lyon 1. He joined this university in September 2009 and is a member of the INRIA RESO Team at the LIP Laboratory. He received his Ph.D. degree in Computer Science from the University Pierre et Marie Curie in 2008, after earning a M.Sc. in Computer Networks from University Pierre et Marie Curie in 2005 and a M.Sc. in Electronics Engineering from ISEP in 2003. In Spring 2009, he was invited to University of California Santa Cruz as a visiting researcher. His research interests include performance evaluation, queueing theory and wireless networks.
}

\profile{Patrick Loiseau}{%
received a M.Sc. degree in physics (2006) and a Ph.D. degree in computer science (2009) from Ecole Normale Sup\'erieure de Lyon (France). He received a M.Sc. degree in mathematics (2010) from UPMC (U. Paris 6 and Ecole Polytechnique). He was a post-doctoral fellow at INRIA Paris-Rocquencourt (2010) and at UC Santa Cruz (2011). He is currently Assistant Professor in the networking and security department at EURECOM (France).
Patrick Loiseau's main research interests are in the areas of probability, statistics and game theory with applications to networks modeling. He is specifically interested in network traffic modeling, performance evaluation, inference of traffic characteristics (sampling), resource pricing and modeling of network security interactions. He has also worked on large deviations with applications to heart-rate modeling.
}


\end{document}